\documentclass[twocolumn,showpacs,preprintnumbers,amsmath,amssymb]{revtex4-1}

\usepackage{graphicx}
\usepackage{dcolumn}
\usepackage{bm}
\usepackage{epstopdf}
\usepackage{amssymb}
\usepackage{verbatim}
\usepackage{epsfig}
\usepackage{tikz}        
\usepackage{etoolbox}

\usepackage{lineno}
\usepackage[caption=false]{subfig} 

\usepackage{mwe}
\usepackage{subfig}

\begin{document}

\newrobustcmd*{\mydiamond}[1]{\tikz{\filldraw[black,fill=#1] 
(0,0) -- (0.2cm,0.2cm) -- (0.4cm,0) -- (0.2cm,-0.2cm) -- (0,0);}}

\newrobustcmd*{\mytriangleright}[1]{\tikz{\filldraw[black,fill=#1] (0,0.2cm) -- 
(0.3cm,0) -- (0,-0.2cm);}}


\title{The thermal conductivity of ``pastas'' in neutron star matter} 
%

 
  \author{C.O.~Dorso}%
  \affiliation{Departamento de F\'\i sica, Facultad de Ciencias Exactas y 
Naturales, Universidad de Buenos Aires,
 Pabell\'on I, Ciudad Universitaria, 1428 Buenos Aires, Argentina.}
  \affiliation{Instituto de F\'\i sica de Buenos Aires, Pabell\'on I, 
Ciudad Universitaria, 1428 Buenos Aires, Argentina.}

 \email{codorso@df.uba.ar}
 
   \author{Jonathan~Dunn}
  \affiliation{School of Materials Engineering, Purdue University, Neil 
Armstrong Hall of Engineering, 701 West Stadium Avenue, West Lafayette, Indiana, 
47907, USA.}

  \author{G.A.~Frank}
  \affiliation{Unidad de Investigaci\'on y Desarrollo de las 
Ingenier\'\i as, Universidad Tecnol\'ogica Nacional, Facultad Regional Buenos 
Aires, Av. Medrano 951, 1179 Buenos Aires, Argentina.}
 
  \author{Alejandro~Strachan}
  \affiliation{School of Materials Engineering, Purdue University, Neil 
Armstrong Hall of Engineering, 701 West Stadium Avenue, West Lafayette, Indiana, 
47907, USA.}

\date{\today}

\begin{abstract}
This investigation explores the phononic thermal conductivity 
of nuclear star matter as it undergoes the ``topological'' transition to the 
``pasta'' regime, and further down to the solid-liquid phase transition. The 
study was carried out using molecular dynamics simulations with nuclear 
potentials embedded in an effective 
(\textit{i.e.} Thomas-Fermi) Coulomb potential. The thermal conductivity 
experiences a dramatic change within a narrow temperature interval around 
$T\simeq 1\,$MeV. This change accomplishes the ``pasta'' breakdown during a 
heating process. The thermal conductivity by flipping protons' 
or neutrons' velocity further shows a decoupling for asymmetric nuclear star 
matter.  

\end{abstract}


\maketitle

\section{\label{introduction}Introduction}

Neutron stars are expected to exist at temperatures as high as $T\sim 9\,$MeV, 
although a fast cooling process takes place soon after their birth. This 
process is caused by neutrino emission and determines the structural evolution 
of the star (see Ref.~\cite{dorso2017} for an insight on the associated 
neutrino 
opacity). The thermal conductivity, $\kappa$, and viscosity, $\eta$, of neutron 
star matter (NSM) are hypothesized to determine the state of the star 
(specially, its crust) \cite{flowers}.   \\

The crust exhibits nucleon densities from $10^{-11}\,$fm$^{-3}$ to  
$0.1\,$fm$^{-3}$ \cite{shternin2007,nandi2018b}, and its inner region 
(beyond $10^{-4}\,$fm$^{-3}$) exhibits complex structural transformations, 
depending 
on the nucleon density. Lower densities result in nuclei embedded in an
electron and neutron gas. Densities above $\rho\sim 0.01\,$fm$^{-3}$, however, 
may lead to complex structures, denoted \textit{pastas} 
\cite{dorso2012a,dorso2012b,horowitz2008,horowitz2016,nandi2018b}. \\ 

The thermal conductivity across the ``inner crust'' is expected to be 
sensitive to temperature and the fraction of species  
\cite{horowitz2008,horowitz2016}. Although it was first accepted for the 
protons 
and neutrons conductivity ($\kappa_p$ and $\kappa_n$, respectively) to be 
negligible with respect to that of electrons \cite{flowers}, researchers 
pointed 
out that $\kappa_p$ and $\kappa_n$ can actually influence the thermal 
relaxation of a neutron star 
\cite{jones2004,shternin2007,horowitz2008,baldo2013,horowitz2016}. \\

Clearly $\kappa_n$ becomes increasingly relevant as the neutron star matter 
becomes 
neutron-rich \cite{flowers}. In addition, other phenomena, such as strong 
magnetic fields, may reduce the electrons' contribution \cite{horowitz2008}, 
enhancing not only the  $\kappa_n$ contribution but also the $\kappa_p$ 
contribution. The later has been shown to produce considerable reduction of 
kinetic coefficients, even at small proton concetrations
(\textit{i.e.} $\beta$-stable matter) \cite{baldo2013}. \\

The effect of nucleon thermal conductivity on the late-time cooling of neutron 
stars is 
somewhat controversial throughout the literature. The observed late-time 
cooling of neutron stars (e.g., MXB 1659-29) was shown to be consistent with 
low 
thermal conductivities \cite{reddy2016}. However, the existence of 
``impurities'' in the \textit{pasta} environment had to be postulated in order 
to match the right cooling curve \cite{horowitz2016}. Also, the finding of 
``spiral defects''  within this context was considered as an additional source 
of electron scattering \cite{reddy2016,horowitz2015}. \\

Regardless of the existence of defects in the \textit{pasta} 
environment, the true ``effective'' conductivity remains rather 
uncertain. Variations of an order of magnitude may be expected due to the 
alignment of the \textit{pasta} structures with respect to the radial axis of 
the star \cite{horowitz2016}. Randomly oriented ``\textit{pasta} slabs'' 
may reduce the conductivity by 37\%, according to molecular dynamics 
simulations 
reported in Ref.~\cite{horowitz2016}. \\

Some criticism on these conductivity estimates arose in recent years. 
Researchers observe that the use of different potentials may lead to variations 
of one order of magnitude of the nucleon contribution to transport coefficients 
\cite{baldo2017}. The formalism of the ``impurity'' parameters has also been 
questioned, arguing the lack of meaning of (high) impurity fractions out of the 
context of uniform crystals \cite{nandi2018a}.  Current simulations, 
however, are not able to predict more convincing $\kappa$ values for matching 
the late cooling of MXB1659-29 \cite{nandi2018a}. Other phenomena, related to 
the electrical conductivities, may also fail to explain the absence of X-ray 
pulsars with periods larger than 12~s \cite{nandi2018a}.\\

Our concern is placed on the ``inner'' crust situation at low 
to moderate temperatures. The \textit{pasta} phase is expected to dominate 
the topological scene at the sub-saturation densities \cite{dorso2012a,Dorso1}. 
This topological regime has already been studied in the context of the equation 
of state (EoS) \cite{Dorso2,Dorso1,dorso2012a}. But research on the thermal 
conductivity still focuses on the energy carried by electrons, disregarding  
the energy flux due to collisions between nuclear species 
\cite{horowitz2008,horowitz2016,nandi2018a}. Other research areas have 
insisted on the role of the non-electronic heat carriers in lattice-like or 
liquid-like systems \cite{kim_2014,stanley_2011}.  \\

We will focus on the heat conduction due to nucleons in the 
\textit{pasta} regime. We presume that topological structures may enhance or 
hinder the energy transport due to nuclear carriers, as first observed in 
Ref.~\cite{dunn}. We will consider, however, that nucleons are embedded in an 
electron gas environment, in order to accomplish a charge-neutral system of 
nucleons and electrons. The term ``thermal conductivity'' in this context means 
the ``phononic'' or ``lattice'' contribution to the thermal conductivity. No 
further mention to the electron contribution will be done. \\

The paper is organized as follows. Section~\ref{background} summarizes 
the theoretical background for the thermal conductivity $\kappa$ in the context 
of the molecular dynamics model (MD). Section~\ref{simulations} explains the 
preparations for measuring $\kappa$ within the \textit{pasta} scenario. The 
corresponding results are exhibited in Section~\ref{results}. For clarity 
reasons, we separated the analysis into symmetric and non-symmetric matter. Our 
conclusions are presented in Section~\ref{conclusions}. \\

\section{\label{background}Background}

We use classical molecular dynamics (CMD) to characterize the thermal transport 
of nuclear star matter.  This approach naturally drives the system to its free 
energy minima within a very complex energy landscape given by the 
inter-particle 
interactions and boundary conditions. Literature results  on the validity of 
this approach can be found in 
Refs.~\cite{Dorso1,dorso2012a,dorso2012b,dorso2014a,lopez2014b}. 

\subsection{\label{potentials} The potentials}

Nuclear matter is considered as a three particle system composed of protons, 
neutrons and electrons. The latter, however, is envisaged as a gas that actually
introduces a screening effect on the Coulomb potential between protons. 
The potentials for  neutron-proton (\textit{np}), neutron-neutron (\textit{nn}) 
and proton-proton (\textit{pp}) interactions were first set by Pandharipande to 
attain a binding energy at the saturation density of 
$E(\rho_0)=-16\,$MeV/nucleon and a compressibility of 250$\,$MeV. The electrons 
screening effect was later introduced through a Thomas-Fermi potential with an 
``effective'' screening length $\lambda$ of $20\,$fm \cite{dorso2014b}. The 
whole set reads as follows\\

\begin{equation}
\begin{array}{rcl}
        V_{np}(r) & = &
\displaystyle\frac{V_{r}}{r}e^{-\mu_{r}r}-\displaystyle\frac{V_{r}}{r_c}e^{-\mu_
{ r } r_ { c } } -\displaystyle\frac{V_ { a }}{r}
e^{-\mu_{a}r}+\displaystyle\frac{V_{a}}{r_{c}}e^{-\mu_{a}r_{c}}\\
       & & \\
       V_{nn}(r) & = &
\displaystyle\frac{V_{0}}{r}e^{-\mu_{0}r}-\displaystyle\frac{V_{0}}{r_{c}}e^{
-\mu
_{0}r_{c}} \\
       & & \\
      V_{pp}(r) & = & V_{nn}(r)  + 
\displaystyle\frac{q^2}{r}e^{-r/\lambda}-\displaystyle\frac{q^2}{r_c'}e^{
-r_c'/\lambda }\\
       \end{array}\label{potentials}
\end{equation}

\noindent where $q$ is the electron charge and $r_c$, $r_c'$ are the cutoff 
distances for the Pandharipande an Thomas-Fermi potentials, respectively. The 
value for the parameters appearing in Eq.~\ref{potentials} can be seen in 
Table~\ref{table_parameter}.\\ 

\begin{table}
{\begin{tabular}{l @{\hspace{15mm}}@{\hspace{6mm}} r @{\hspace{30mm}} l}
\hline
      Parameter & $\ $Value  & \multicolumn{1}{c}{Units} \\
\hline
$V_r$   &  3097.0  & MeV \\
$V_a$   &  2696.0  & MeV\\
$V_0$   &  379.5   & MeV\\
$\mu_r$ &  1.648 & fm$^{-1}$ \\
$\mu_a$ &  1.528 & fm$^{-1}$ \\
$\mu_0$ &  1.628 & fm$^{-1}$ \\
$\lambda$ & 10   & fm \\
$r_c$   &  5.4 & fm \\
$r_c'$  &  20  & fm \\
\hline
\end{tabular}
}
\caption{Parameter set for the CMD computations (New Medium model).  }
\label{table_parameter}
\end{table}

\subsection{\label{conductivity} The thermal conductivity}

The thermal conductivity $\mathbf{\kappa}$ corresponds to the set of 
transport coefficients relating the heat flux (\textit{i.e.} energy flux 
$\mathbf{J}$) to the temperature gradient $\nabla T$, through the 
(phenomenological) Fourier law

\begin{equation}
 \mathbf{J}(t)=-\mathbf { \kappa }\,\nabla T\label{eqn:def_kappa}
\end{equation}

\noindent where $\mathbf{\kappa}$ corresponds to a second rank $3\times 3$ 
tensor for 
non-isotropic matter. Notice that the constitutive relation 
(\ref{eqn:def_kappa}) is intended as a ``macroscopic'' one, whenever matter is 
considered as a continuum. The energy flux $\mathbf{J}$ represents a somewhat 
``mean'' flux density $\langle\mathbf{j}\rangle$ transported across a small 
volume $\mathcal{V}$ (that is, 
$\mathbf{J}=\langle\mathbf{j}\rangle.\mathcal{V}$). 
The calculation of thermal transport properties  from atomistic simulations is 
well established.\cite{muller,zhou2007phonon,dunn2016role,lin2013thermal} \\

We will only consider those situations where  $\mathbf{J}$ and $\nabla T$ are 
collinear (say, for example, along the $\hat{z}$ axis) and use a non-equilibrium
method for computing the thermal conductivity $\kappa_z$ is proposed by 
M\"uller-Plathe (see 
Ref.~\cite{muller}) from the average heat flux and temperature gradient.

\begin{equation}
 \kappa_z=-\lim_{t\rightarrow\infty}\displaystyle\frac{\langle 
J_z\rangle}{\langle\partial T/\partial z\rangle}\label{eqn:kappa_lim}
\end{equation}

If the medium is isotropic, common practice sets the mean 
thermal conductivity as $(k_x+k_y+k_z)/3$. \\

Notice that the linear nature of Eqs.~(\ref{eqn:def_kappa}) and 
(\ref{eqn:kappa_lim}) requires relatively small temperature gradients.  \\

\begin{figure}[htbp!]
\includegraphics[width=0.9\columnwidth]
{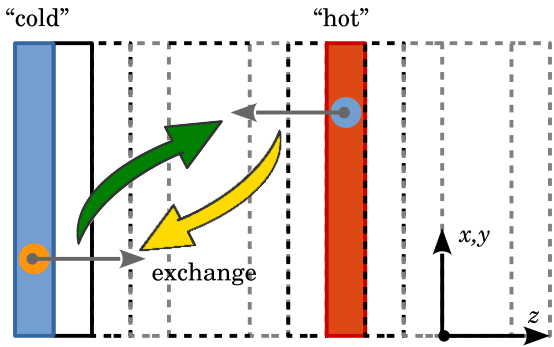}
\caption{\label{scheme} (On-line color only) Schematic representation of the 
M\"uller-Plathe procedure. The blue and red bins correspond the the 
``cold'' and ``hot'' slabs, respectively. The horizontal flat arrows stand for 
the particles velocity. The curved arrows (green and yellow, respectively) 
represent the velocity exchange process. } 
\end{figure}

In a nutshell, the M\"uller-Plathe procedure \cite{muller} generates a heat
flux of known magnitude and the temperature gradient is obtained as local
averages of the kinetic energy. The system is divided in thin bins along the 
the 
heat flux direction (see Fig.~\ref{scheme} for  details); the first slab is 
labeled as the ``cold'' slab, while the slab in the middle is labeled as 
``hot''. A heat flux is generated by exchanging the velocities of two particles 
(with the same mass), the hottest particle in the ``cold'' bin and the coldest 
one in the ``hot'' bin (see Fig.~\ref{scheme}). Thus, the system is 
(artificially) driven out of equilibrium, and a heat flux $\mathbf{J}$ develops 
through the system of interest in the opposite direction for the equilibrium 
restoration. This flux is expected to reach the stationary state if the 
exchanging rate is held regularly for a long time.   \\

In order to generate an external heat flux, particle velocity exchanges are
performed periodically during the MD simulation.  Recall that the species 
themselves are not exchanged, but only the velocities. Thus, the ``pumping'' 
process only transports kinetic energy (for particles with the same mass). This 
procedure conserves total energy and linear momentum. \\ 

The heat flux introduced by the velocity exchange is hard to compute from 
dynamical magnitudes. The computation from the net transported (kinetic) energy 
is somewhat easier since

\begin{equation}
 \langle J_z\rangle+\displaystyle\frac{1}{2A_{xy}}\,\bigg[ 
\displaystyle\frac{1}{\tau}\sum_{n=1}^T\displaystyle\frac{1}{2}m(v_{h}^{2}-
v_{c}^{2})\bigg] =0
\label{eqn:J_transported}
\end{equation}

\noindent where the expression between the square brackets represents the mean 
(kinetic) energy exchanged during the time period $\tau$. $v_h$ and $v_c$ 
refers to the velocities of the hot particle and cold particle, respectively.  
The factor $2A_{xy}$ corresponds to the cross section of the slabs (two faces). 
 \\

The temperature profile is obtained by computing the local (kinetic) 
temperature for each slab. Once steady state is reached, the temperature 
profile 
is expected to be linear away from the cold and hot bins where velocities are 
exchanged, 
provided the heat flux remains small. Further details can be found in 
Ref.~\cite{muller}. 
 \\

We stress the fact that the balance condition (\ref{eqn:J_transported}) 
links the heat flux $\mathbf{J}$ to the (artificial) kinetic energy 
transportation introduced by the M\"uller-Plathe procedure. The velocity 
exchange is not restricted to pairs of similar particles, but also across 
species (with the same mass). Therefore, the procedure enables the computation 
of the thermal conductivity for the set of \textit{all} the nucleons, or for 
the set of protons and neutrons separately. The meaning of either coefficients, 
though, will be quite different.  \\

\section{\label{simulations}Simulations}

The MD simulations were performed using the Verlet integration scheme 
(with periodic boundary conditions). The Nos\'e-Hoover thermostat was used to 
drive the system to the desired temperature with a coupling constant of 
10 (time units). All simulations were performed with the LAMMPS simulation code 
running on GPU's (Graphic Processing Units) \cite{plimpton}. \\

We were able to simulate around 100,000 nucleons in the primary cubic cell. 
However, the \textit{pasta} topology became so complex that the resulting 
thermal conductivity represented that of a complex structure. Thus, we 
restricted our study to 
the most elementary topologies (say, straight \textit{lasagnas} or 
\textit{spaghettis}) accomplished by a system size of 4,000 nucleons.  \\

At a first instance, the system was cooled from $T=4\,$MeV down to the 
solid (pasta) state (say, $T=0.1\,$MeV). The density $\rho$ ranged from 
$0.03\,$fm$^{-3}$ to $0.05\,$fm$^{-3}$. Nice \textit{lasagnas} or 
\textit{spaghettis} resulted after the cooling, although not completely aligned 
to the canonical axes (see below). In order to improve the alignment, we 
softened the \textit{pasta} by raising the bath temperature to $0.8-1.2\,$MeV,  
and then, we performed the corresponding transformations. The \textit{pasta} 
was 
finally cooled back to $0.1\,$MeV. \\

At a second instance, the bath temperature was increases from $0.1\,$MeV to 
$2.1\,$MeV, while the nucleons' positions and velocities were recorded at
regular time intervals. The recorded configurations were set as the initial 
conditions for the thermal conductivity measurements. \\

The thermal conductivities reported in Section~\ref{results} correspond to 
those obtained following the M\"uller-Plathe procedure (see 
Ref.~\cite{muller}). 
 Data was collected after a steady state was reached from each initial 
condition (within the fluctuations typical of small systems). Recall that the 
M\"uller-Plathe 
procedure is known to attain a precision of 10\% (on a system of 2600 
Lennard-Jones particles \cite{muller}).  \\  

The M\"uller-Plathe procedure requires the binning of the primary cell, in 
order to compute the temperature gradient across the bins (that is, along the 
heating flux direction). This is why we demanded a proper alignment of the 
\textit{pasta} with respect to the canonical axes. We set the number of bins 
to 20. \\

For each pasta topology we computed  two values of the thermal conductivity:  
``parallel'' and ``transverse''. The former corresponds to the heat flux along 
the \textit{pasta}. The latter corresponds to the heat flux  running across the 
\textit{pasta}. Each measurement were computed separately.   \\ 

Whatever the heat flux direction, we computed the thermal conductivity 
by either flipping the protons' and neutrons' velocity 
separately, and \textit{all} the nucleons regardless of 
their nature. The former means that only one specie contributes 
to the velocity exchange in Eq.~(\ref{eqn:J_transported}). This distinction 
became very useful when analyzing non-symmetric matter (see 
Section~\ref{asymmetry}). \\  

\begin{figure}[htbp!]
\includegraphics[width=\columnwidth]
{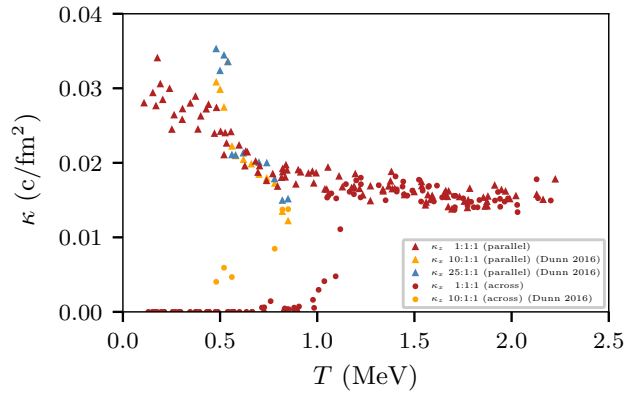}
\caption{\label{kappa_all_raw_data} (On-line color only) Thermal conductivity 
$\kappa$ vs. temperature. The red symbols correspond to our simulations during 
the heating evolution after the \textit{lasagna} was established. The orange 
and blue symbols correspond to literature data. All data points correspond to a 
system density of $\rho=0.05$ ($x=0.5$) and computations were done on 
\textit{all} the nucleons. The aspect ratio of simulation cells is indicated in 
the insert.   } 
\end{figure}

We made a preliminary check of our result with the few data reported in 
literature \cite{dunn}. Fig.~\ref{kappa_all_raw_data} shows our output for an  
1:1:1 simulation cell and the reported data for larger cells. The thermal 
conductivity was computed over \textit{all} the nucleons arranged in a 
\textit{lasagna}-like topology. \\

Notice from Fig.~\ref{kappa_all_raw_data} that our preliminary results 
match the ones reported in Ref.~\cite{dunn} for the ``parallel'' measurement 
(triangular symbols in Fig.~\ref{kappa_all_raw_data}). Our ``across'' 
measurements, however, do not completely agree with Ref.~\cite{dunn}. But since 
the number of slabs (and the space binning) in Ref.~\cite{dunn} may be 
different from ours, no definite conclusions can be drawn about the mismatch.
Nevertheless, both data sets exhibit a similar qualitative behavior. 
Recall that our objective is understanding the influence of \textit{pasta} 
structures on the energy transport.    \\

\section{\label{results}Results}

\subsection{\label{caloric_curve} Symmetric matter}

\subsubsection{\label{caloric_curve} The caloric curve}

\begin{figure}[htbp!]
\includegraphics[width=\columnwidth]
{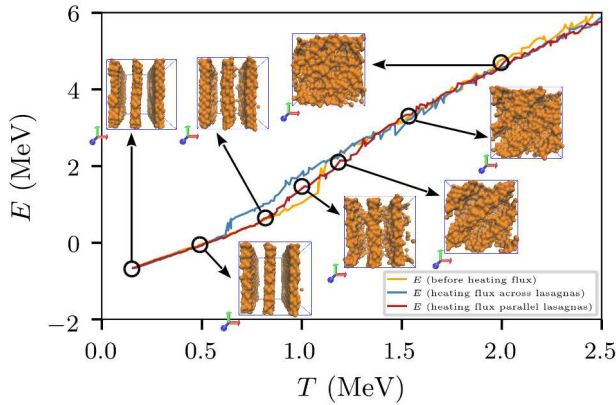}
\caption{\label{heating_energy} (On-line color only) Internal energy vs. 
temperature during the heating evolution after a \textit{lasagna} was 
established. The system density was $\rho=0.05$ and the proton ratio was 
$x=0.5$. The surface plots show only the protons' topology. Three situations 
are represented: (a) The energy evolution just before a heat flux was 
induced (orange line). (b) The energy evolution after the a heat flux was 
established across the \textit{lasagnas} (blue line). (c) The energy evolution 
after the a heat flux was established parallel to the \textit{lasagnas} 
(red line). } 
\end{figure}

As mentioned in Section \ref{simulations}, we focused on relatively small 
cells housing 4,000 nucleons (with periodic boundary conditions). The 
simulation cell was previously cooled according to the protocol detailed in 
Section \ref{simulations}. Fig~\ref{heating_energy} (orange line) shows the 
energy (per nucleon) during heating following this initial annealing.\\

The M\"uller-Plathe procedure introduces a periodical velocity exchange between 
nucleons, as explained in Section~\ref{conductivity}. This kind of energy 
exchange induces a heat flux across the cell, while the corresponding 
temperature gradient appears along the same flux direction. This results in a 
perturbed system due to this (externally) induced heat flux. 
Fig~\ref{heating_energy} shows the internal energy (per nucleon) when 
the heat flux is parallel (red line) and transverse (blue line) to the 
\textit{lasagna}. Comparisons can be made with respect to the energy of the 
unperturbed system. The inserted snapshots correspond to the protons' 
distribution after a parallel heat flux was established  (see caption for 
details). \\

The internal energy prior to any measurement perturbation (say, before 
introducing a heat flux in a specific direction) exhibits a sharp ``jump'' 
at $T\simeq 1.25\,$MeV (see orange line in Fig.~\ref{heating_energy}). This 
jump corresponds to the ``pasta'' breakdown (\textit{i.e.} the topological 
transition \cite{Dorso1}). Interestingly, Fig.~\ref{heating_energy} also 
shows that the energy-temperature relationship changes when a heat flux (and a 
temperature gradient) is introduced,  using the M\"uller-Plathe procedure in 
our case. The snapshots in Fig.~\ref{heating_energy} capture the relation, at 
least qualitative, that exists between the energy profile and the 
changes in the system topology.  \\ 

\begin{figure}[htbp!]
\includegraphics[width=\columnwidth]
{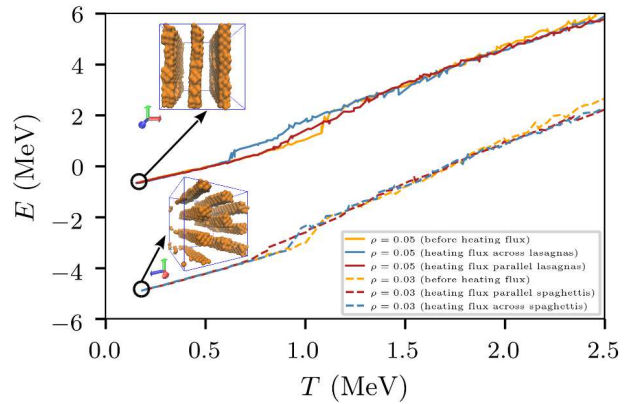}
\caption{\label{heating_energy_2} (On-line color only) Internal energy vs. 
temperature during the heating evolution after the \textit{pasta} was 
established. Two different pastas are represented: a spaghetti-like system 
($\rho=0.03$ and $x=0.5$), and a lasagna-like system ($\rho=0.05$ and $x=0.5$). 
The former corresponds to the dashed lines, while the later corresponds to the 
continuous lines. } 
\end{figure}

A comparison between the internal energy curve of \textit{spaghettis} and 
\textit{lasagnas} can be performed from Fig.~\ref{heating_energy_2}. Notice 
that 
both curves are essentially the same, although the \textit{lasagna} profile 
appears shifted upward with respect to the \textit{spaghetti} one. The energy 
``jump'' is also present near $T\simeq 1.25\,$MeV for the spaghettis, achieving 
the corresponding topological breakdown. The M\"uller-Plathe perturbation can 
further be seen.  \\

\subsubsection{\label{thermal_conductivity} The thermal conductivity $\kappa$}

\begin{figure*}[htbp!]
\subfloat[wide view]{\includegraphics[width=\columnwidth]
{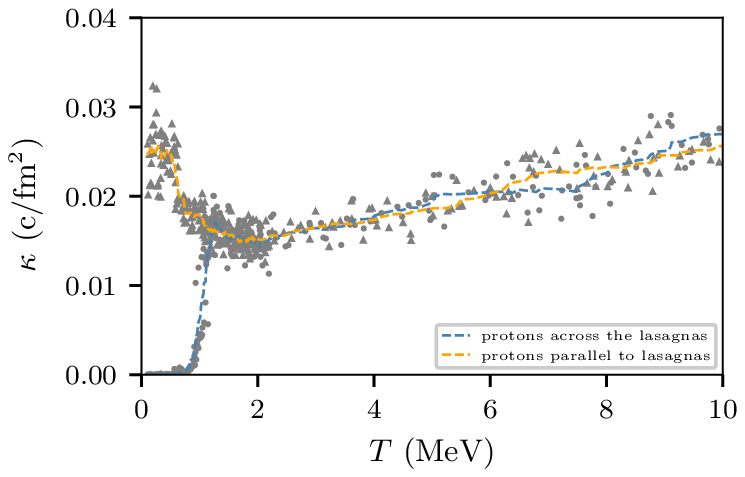}\label{kappa_all_vs_protons_1}}
\subfloat[detailed view]{\includegraphics[width=\columnwidth]
{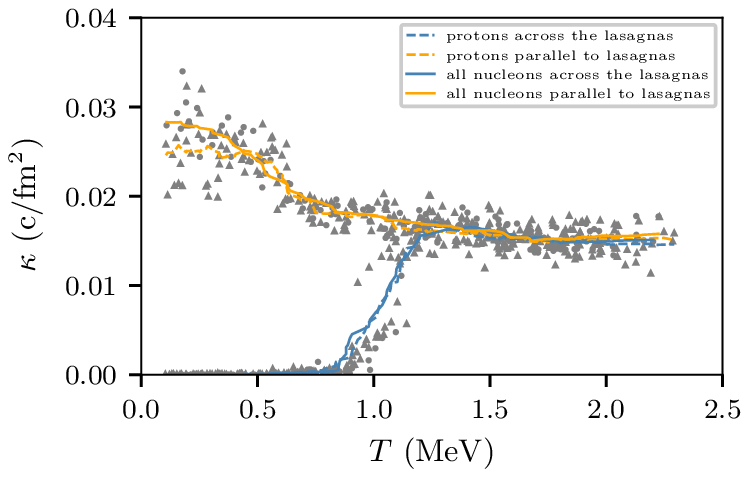}\label{kappa_all_vs_protons_2}}
\caption[width=0.47\columnwidth]{\label{kappa_all_vs_protons} (On-line color 
only) Thermal conductivity $\kappa$ vs. temperature during the heating 
evolution 
after the \textit{lasagna} was established (see Fig.~\ref{heating_energy} for 
the corresponding snapshots). The system density was $\rho=0.05$ ($x=0.5$). The 
smoothening was done following a moving average procedure of $\pm10$ points. 
The rounded gray points correspond to the raw data obtained over \textit{all} 
the nucleons. The triangular gray points correspond to the raw data obtained 
over \textit{protons} only.  } 
\end{figure*}

We first computed the thermal conductivity for symmetric neutron star matter 
($x=0.5$) as a function of temperature in the $0.1-2.1\,$MeV range. This 
includes the solid-liquid transition ($T\sim 0.5\,$ MeV) and the topological 
transition ($T\sim 1\,$MeV). The computation was carried out in two ways: by 
considering the heat flux due to \textit{all} the nucleons, or, considering 
only \textit{one} kind on nucleons (say, the protons; see 
Section~\ref{simulations} for details). Fig.~\ref{kappa_all_vs_protons_1} shows 
the proton thermal conductivity for a wide temperature range. 
Fig.~\ref{kappa_all_vs_protons_2} shows the details of the smoothed data 
computations obtained for either protons and \textit{all} the nucleons. \\

Two regimes can be distinguished immediately according to 
Fig.~\ref{kappa_all_vs_protons}. The thermal conductivity  exhibits a smooth 
slope above $T\simeq 1.25\,$MeV, while a dramatic change occurs below this 
threshold. The later appears as a ``decoupling'' between the thermal 
conductivity parallel to the \textit{lasagna} ($\kappa_z$) and the one 
orthogonal to this direction ($\kappa_x$). The ``decoupling'' pattern is 
essentially the same whether all the nucleons are considered or only the 
protons 
(for $x=0.5$). \\

It can be noticed in Fig.~\ref{kappa_all_vs_protons} that the conductivity for 
\textit{all} the nucleons appears somewhat shifted up with respect to the 
protons' conductivity. The large fluctuations in the data do not allow a 
definite conclusion on this phenomenon. However, a small bias seems reasonable 
due the (local) density of the considered specie (say, nucleons or protons 
only). An insight to this issue is given at the end of this Section.  \\

\begin{figure}[htbp!]
\includegraphics[width=\columnwidth]
{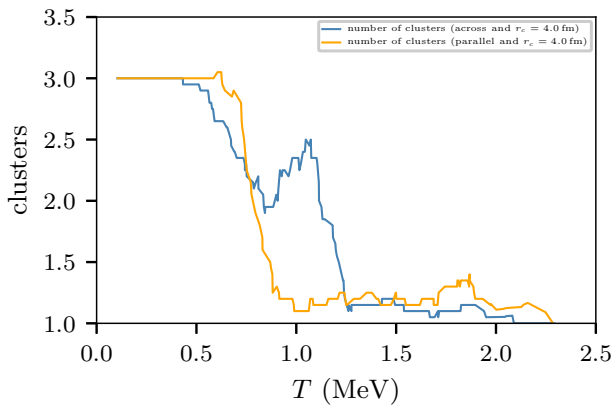}
\caption{\label{clusters_rho05_05_400} (On-line color only) Number of clusters 
in the simulation cell for the lasagna-like structures shown in 
Fig.~\ref{heating_energy}. Data corresponds to two heat flux: parallel or 
across the lasagna structure (see insert for details). The 
smoothening was done following a moving average procedure of $\pm10$ points. 
The cut-off distance was set to $4\,$fm.   } 
\end{figure}

We turn to study the topology of the nucleons in the simulation cell for a 
better understanding of the ``decoupling threshold''. We measured the number of 
clusters (of the 4000 nucleons) within the simulation cell. Only the particles 
in the cell were taken into account while performing the cluster analysis (no 
images included). In fact, the \textit{lassagnas} are infinite in, say the 
$(x,y)$ plane, but the system is discontinuous in the $\hat{z}$ direction.
Fig.~\ref{clusters_rho05_05_400} shows the results for the evolution in 
Fig.~\ref{heating_energy}. The cut-off distance between neighbors belonging to 
the same cluster was set to $r_c=4\,$fm, in order to get exactly three 
clusters at $T=0.1\,$MeV (see Fig.~\ref{heating_energy}). \\

According to Fig.~\ref{clusters_rho05_05_400}, the number of clusters drops off
near $T=1\,$MeV. This corresponds to the \textit{lasagna} breakdown shown in 
Fig.~\ref{heating_energy}. The fluctuations observed at the drop off interval 
(say, $1-1.25\,$MeV) correspond to weak (or temporary) connections between 
slabs during the evolution. These can be noticed in 
Fig.~\ref{kappa_all_vs_protons} by watching the raw data points, although its 
trace is lost after the smoothening procedure. The ``decoupling'' threshold 
is thus associated to this enhanced connectivity among nucleons (meaning the 
pasta breakdown). \\

The vanishing values of the thermal conductivity across a well established 
\textit{lasagna} (say, for $T<1\,$MeV) can be easily explained because of the 
existence of voids between the slabs. The negative slope for the parallel 
$\kappa$ (\textit{i.e.} along the \textit{lasagna}) means that the ``solid 
pasta'' presents an enhanced conductivity with respect to the ``liquid pasta''. 
This behavior is common to other materials. \\   

\begin{figure}[!htbp]
\subfloat[$\rho=0.03$]{\includegraphics[width=0.47\columnwidth]
{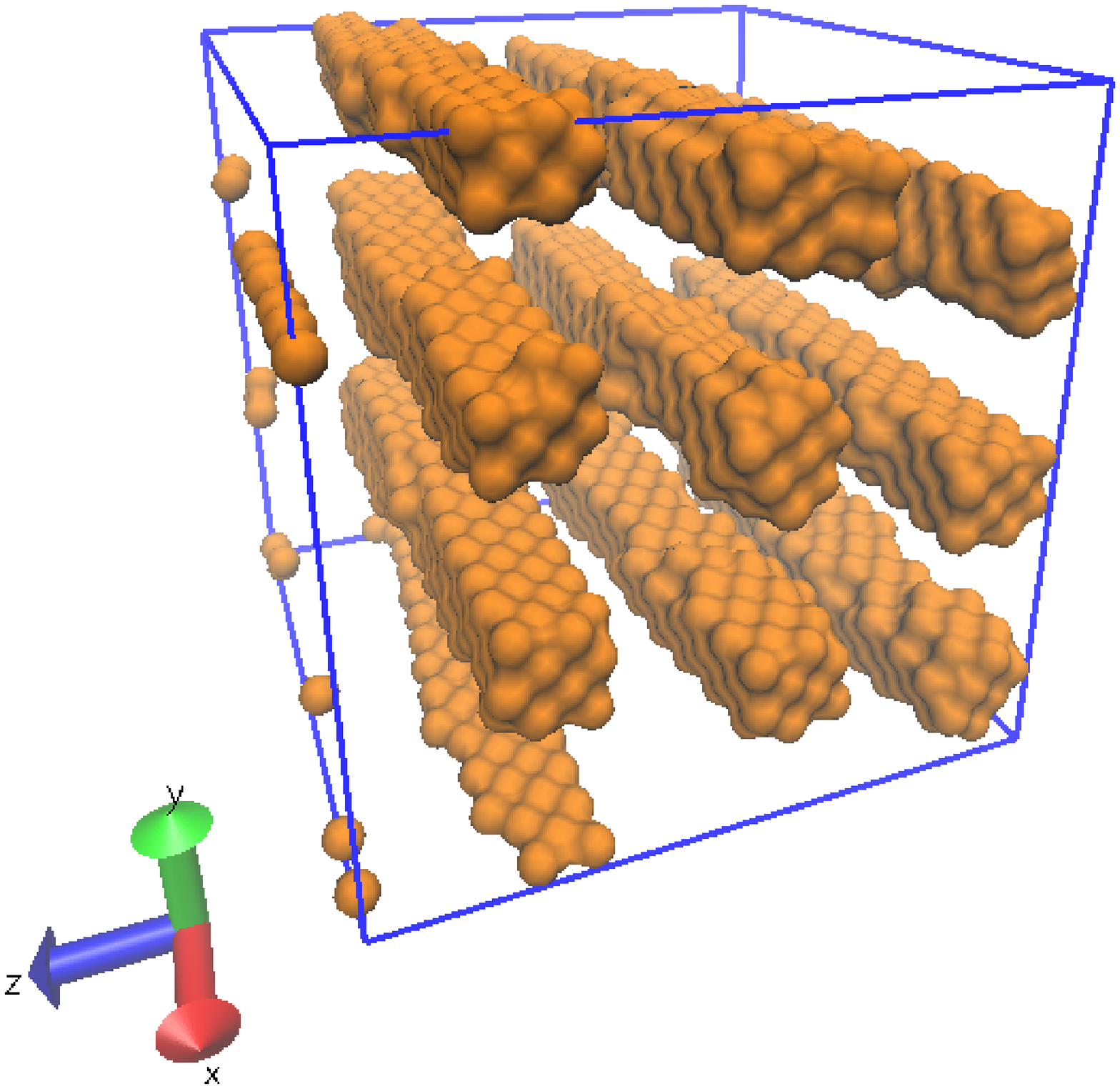}
\label{output_rho03_05}}
\subfloat[$\rho=0.04$]{\includegraphics[width=0.47\columnwidth]
{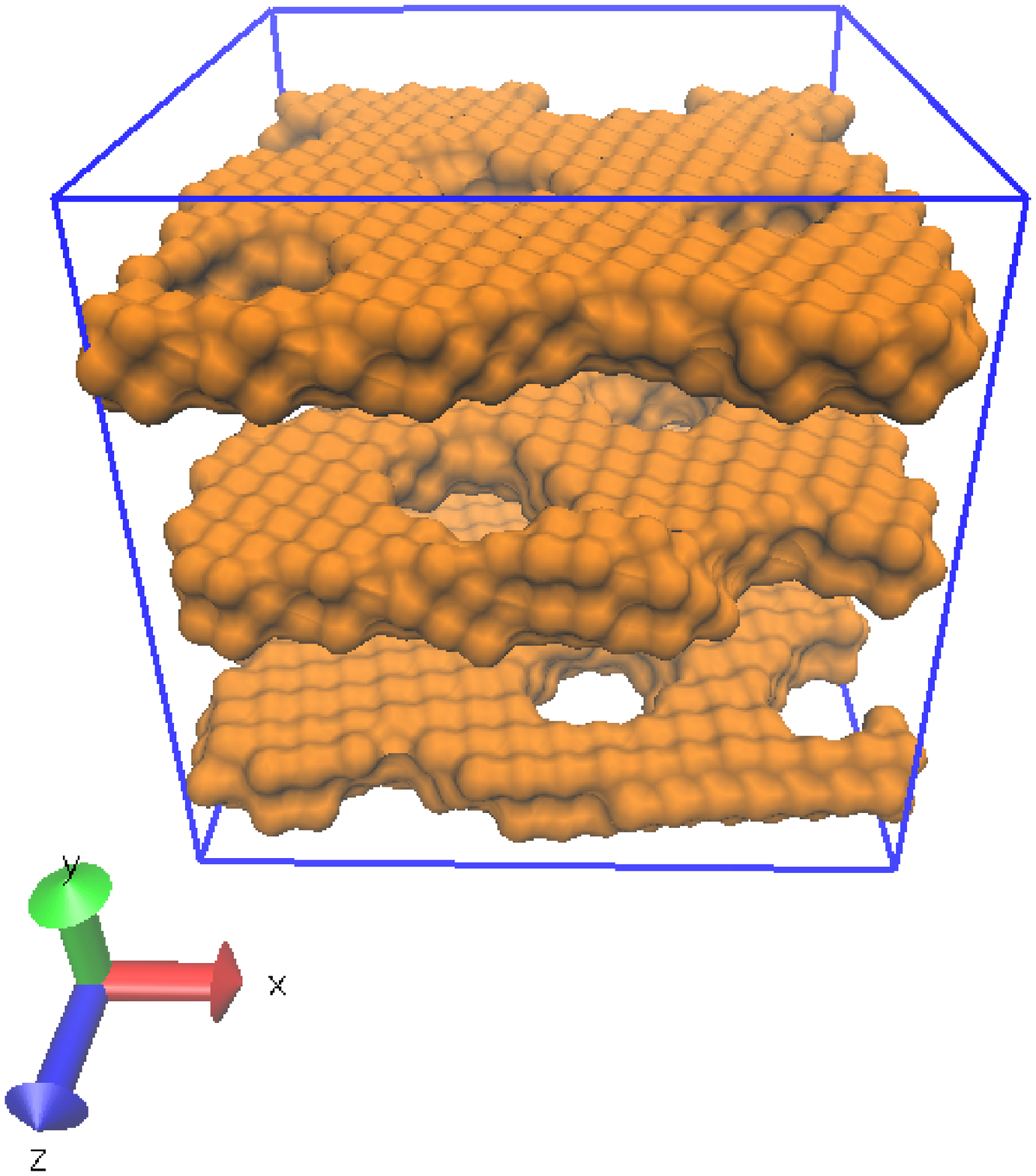}
\label{output_rho04_05}}\\
\subfloat[$\rho=0.045$]{\includegraphics[width=0.47\columnwidth]
{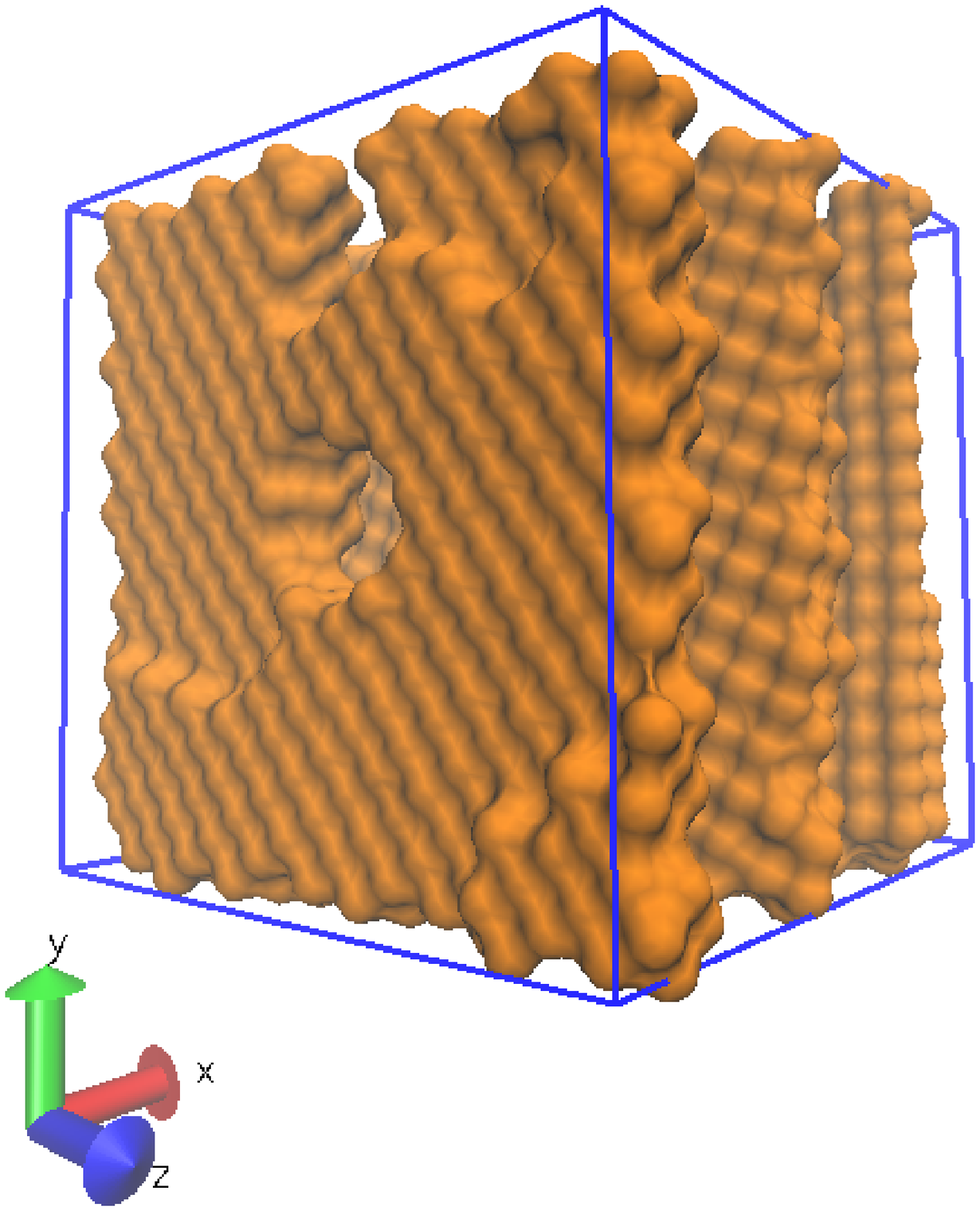}
\label{output_rho045_05}}
\subfloat[$\rho=0.05$]{\includegraphics[width=0.47\columnwidth]
{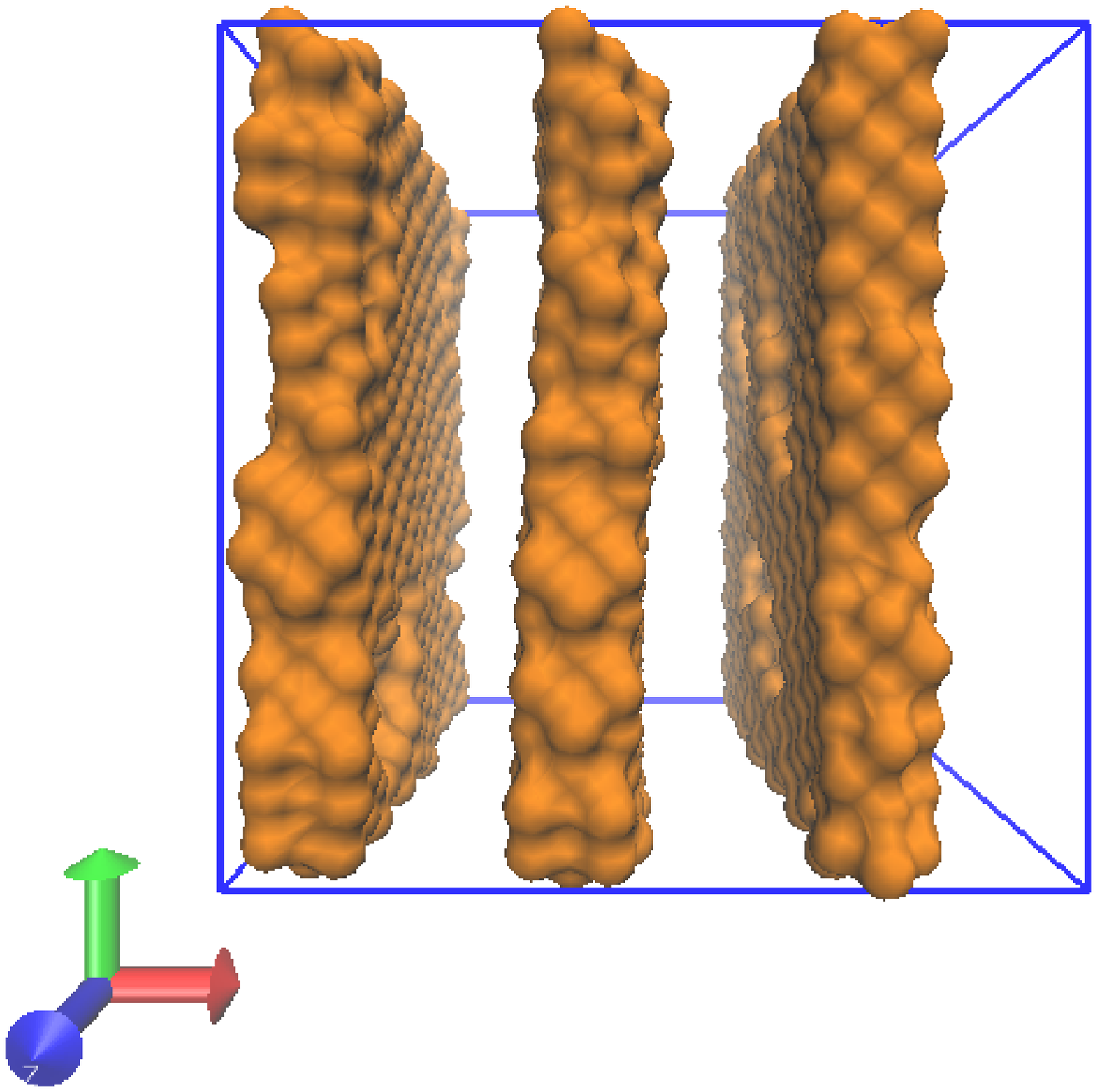}
\label{output_rho045_05}}
\caption[width=0.47\columnwidth]{Surface plots for protons at $T=0.1\,$MeV 
($x=0.5$). The nucleon densities are indicated below each snapshot. } 
\label{snaphots_densities}
\end{figure}

\begin{figure}[htbp!]
\includegraphics[width=\columnwidth]
{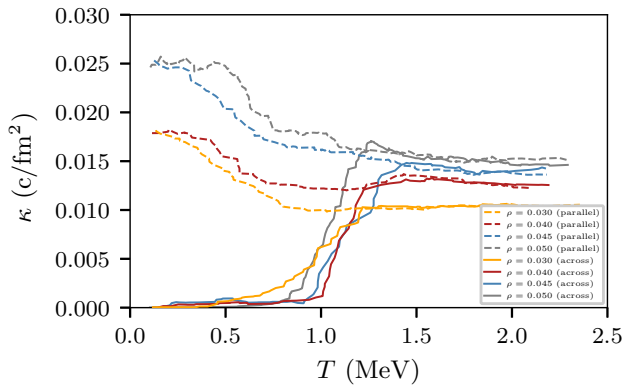}
\caption{\label{kappa_protons_densities} (On-line color only) Proton thermal 
conductivity $\kappa$ vs. temperature for densities in the range 0.03 to 
0.05$\,$fm$^{-3}$ and $x=0.5$ (see insert for details). The smoothening was 
done following a moving average procedure of $\pm10$ points. The dashed lines 
correspond to the thermal conductivity values along (parallel) the pasta 
structure. The continuous lines correspond to the thermal conductivity across 
the pasta structure.  } 
\end{figure}

The slab structure of the \textit{lasagnas} undergoes openings for decreasing 
densities in the simulation cell. Fig.~\ref{snaphots_densities} shows how these 
openings spread over the slabs until the \textit{lasagna} is no longer 
sustainable, moving to the \textit{spaghetti}-like structure (see 
Fig.~\ref{output_rho03_05}). Fig.~\ref{kappa_protons_densities} exhibits the 
corresponding proton thermal conductivity (after the data smoothening). \\

The ``decoupling'' pattern goes through the explored densities (see 
Fig.~\ref{kappa_protons_densities}), including either \textit{spaghettis} or 
\textit{lasagnas}. The ``decoupling threshold'' (say, $T\simeq 1.25\,$MeV) 
remains unchanged (within the current measurement errors). The 
(parallel) thermal conductivity, however, exhibits a density dependency on 
either side of this threshold. According to Fig.~\ref{kappa_protons_densities}, 
the parallel $\kappa$ (protons only) increases for increasing densities all 
along the explored temperatures. The orthogonal $\kappa$ (protons only) meet 
this behavior above the ``decoupling threshold'', that is, after the pasta 
breakdown occurs. \\

Notice that the solid-like state also attains some kind of density dependence 
for $\kappa$ (protons only). The current fluctuations of our measurements does 
not allow to distinguish clearly between the corresponding thermal conductivity 
values at $\rho=0.03$ (\textit{spaghettis}) and $\rho=0.04$ (washed out 
\textit{lasagna}). But Fig.~\ref{kappa_protons_densities} shows fairly 
different values between $\rho=0.03$ and $\rho=0.05$.\\

We may summarize our results as follows. The \textit{pasta} breakdown process 
(during a heating evolution) accomplishes a dramatic change in the thermal 
conductivity of symmetric neutron star matter. For ``cold'' \textit{pastas}, 
the thermal conductivity is only possible along the \textit{pasta} structure, 
attaining a ``decoupling'' between orthogonal directions. The solid state of 
``cold'' \textit{pastas} even enhances the conductivity. But warming the 
\textit{pastas} above the threshold $T\simeq 1.25\,$MeV, breaks down its 
topological structure, connecting regions that were once separated by voids. 
This situation allows heating on any direction, and thus, the thermal 
conductivity switches to an homogeneous (isotropic) value, that may depend on 
the system density. \\

\subsection{\label{asymmetry} Non-symmetric matter}

\begin{figure}[htbp!]
\includegraphics[width=0.8\columnwidth]
{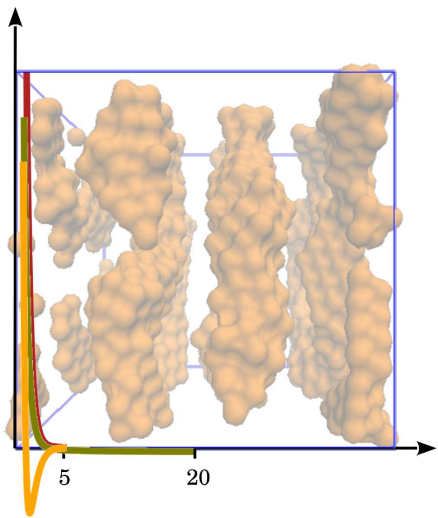}
\caption{\label{potentials_plot} (On-line color only) Potentials 
as a function of distance (fm). The red curve corresponds to $V_{nn}(r)$, the 
orange curve corresponds to $V_{np}(r)$ and the green one to $V_{pp}(r)$ 
(includes the Coulomb contribution. The semi-transparent image in the 
background represents a \textit{lasagna}-like structure for $\rho=0.05$ and 
$x=0.3$ (only protons are represented).   } 
\end{figure}

The next step in the investigation focused on the thermal conductivity behavior 
for proton ratios varying from $x=0.5$ down to $x=0.3$. The nucleons' 
potentials 
remained unchanged, as expressed in Section~\ref{potentials}. 
Fig.~\ref{potentials_plot} shows the corresponding profiles (up-to the cut-off 
distance) in comparison with a \textit{lasagna}-like background. Notice that 
the 
slabs widths do no exceed the cut-off distance, although they look more 
irregular than in the case of symmetric matter. For the sake of robustness, we 
will set the cut-off distance for clusters recognition to $5\,$fm.     \\ 

\subsubsection{\label{asymmetry_caloric} The caloric curve}

\begin{figure}[htbp!]
\includegraphics[width=\columnwidth]
{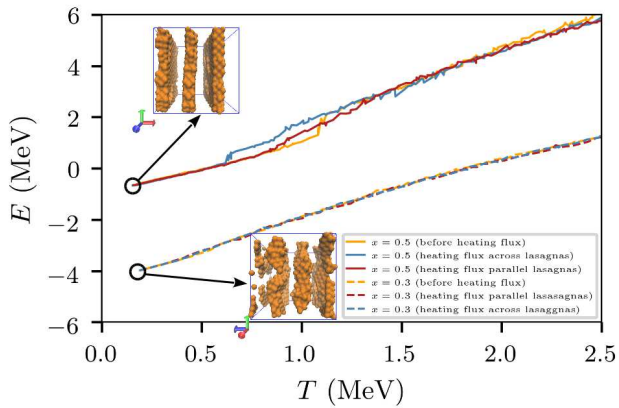}
\caption{\label{heating_energy_3} (On-line color only) Internal energy vs. 
temperature during the heating evolution after the \textit{pasta} was 
established. Two lasagna-like systems are represented ($x=0.3$ and $x=0.5$) 
for the density $\rho=0.05$. The asymmetric situations correspond to the 
dashed lines, while the symmetric ones correspond to the continuous lines. } 
\end{figure}

Fig.~\ref{heating_energy_3} shows the internal energy evolution after the 
(asymmetric) \textit{pasta} was well established for $x=0.3$ \cite{Dorso1}. The 
corresponding energy for the symmetric situation is included for comparison 
reasons. \\ 

The two profiles exhibited in Fig.~\ref{heating_energy_3} are very similar, 
regardless of the obvious overall energy bias. However, a noticeable 
difference can be pointed out: the sharp ``jumps'' observed near $T=1\,$MeV in 
the symmetric situation are actually not visible in the $x=0.3$ situation. \\

\begin{figure}[htbp!]
\includegraphics[width=\columnwidth]
{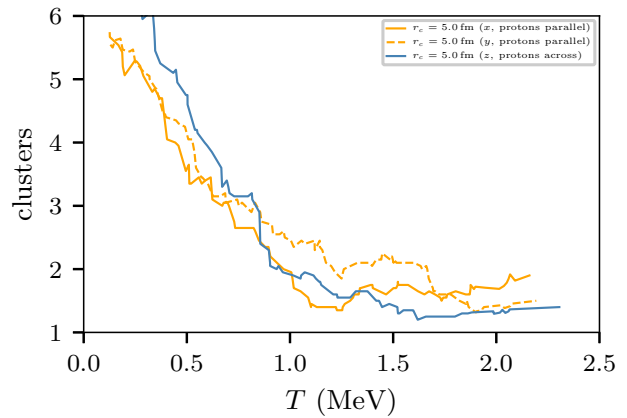}
\caption{\label{clusters_rho05_03_500} (On-line color only) Number of clusters 
in the simulation cell for the lasagna-like structures shown in 
Fig.~\ref{kappa_protons_asymmetric_snapshots}. Data corresponds to two heat 
flux: parallel or across the lasagna structure (see insert for details). The 
smoothening was done following a moving average procedure of $\pm10$ points. 
The cut-off distance was set to $5\,$fm.   } 
\end{figure}

In order to get a more accurate picture of the asymmetric situation ($x=0.3$) 
we computed the proton clusters along the \textit{pasta} regime. The results 
are shown in Fig.~\ref{clusters_rho05_03_500}. The cut-off distance for 
neighboring protons belonging to the same cluster was set to $r_c=5\,$fm. \\

\begin{figure}[!htbp]
\subfloat[non-transparent]{\includegraphics[width=0.51\columnwidth]
{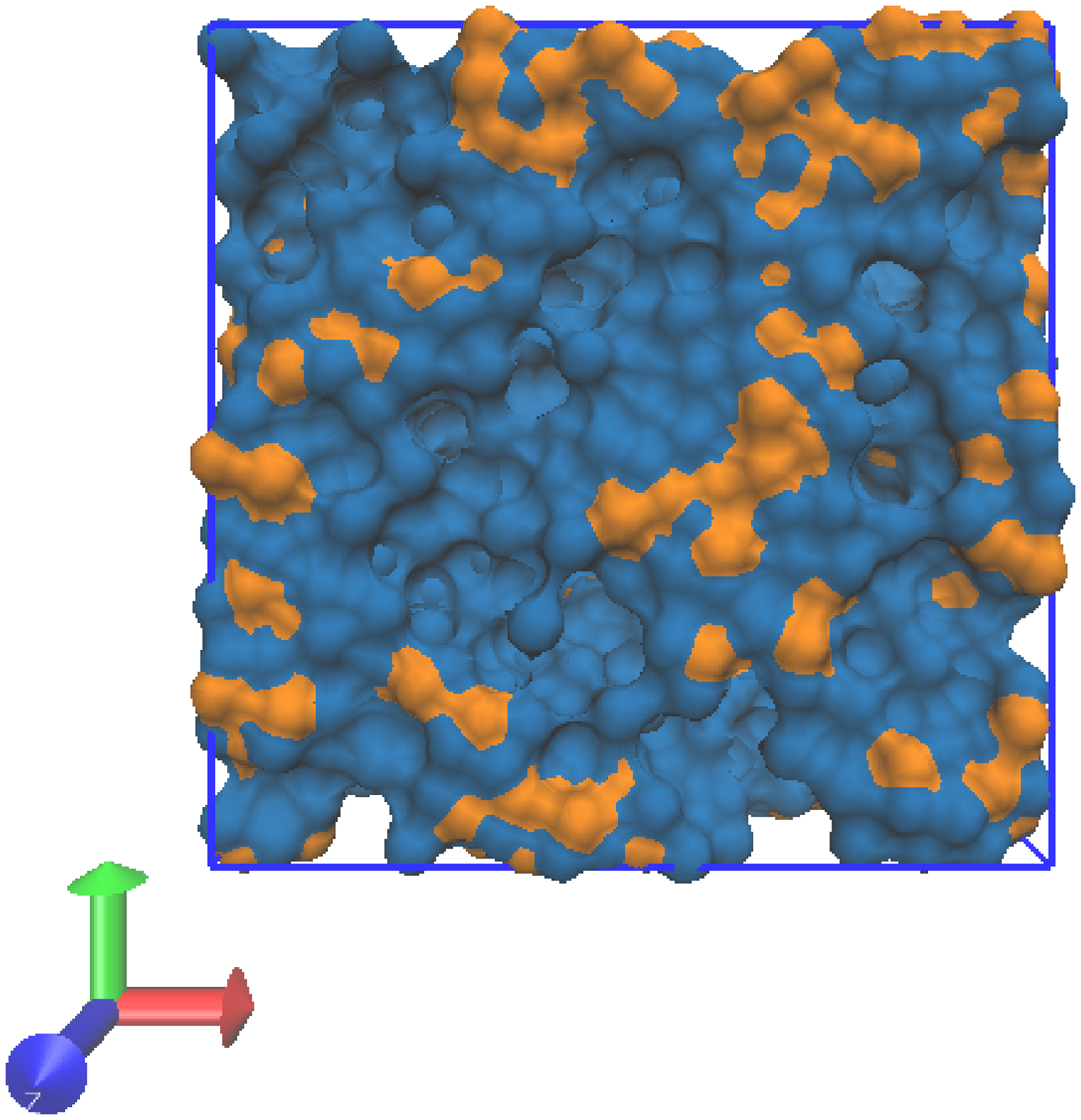}
\label{snaphots_x03_full_07}}
\subfloat[semi-transparent]{\includegraphics[width=0.51\columnwidth]
{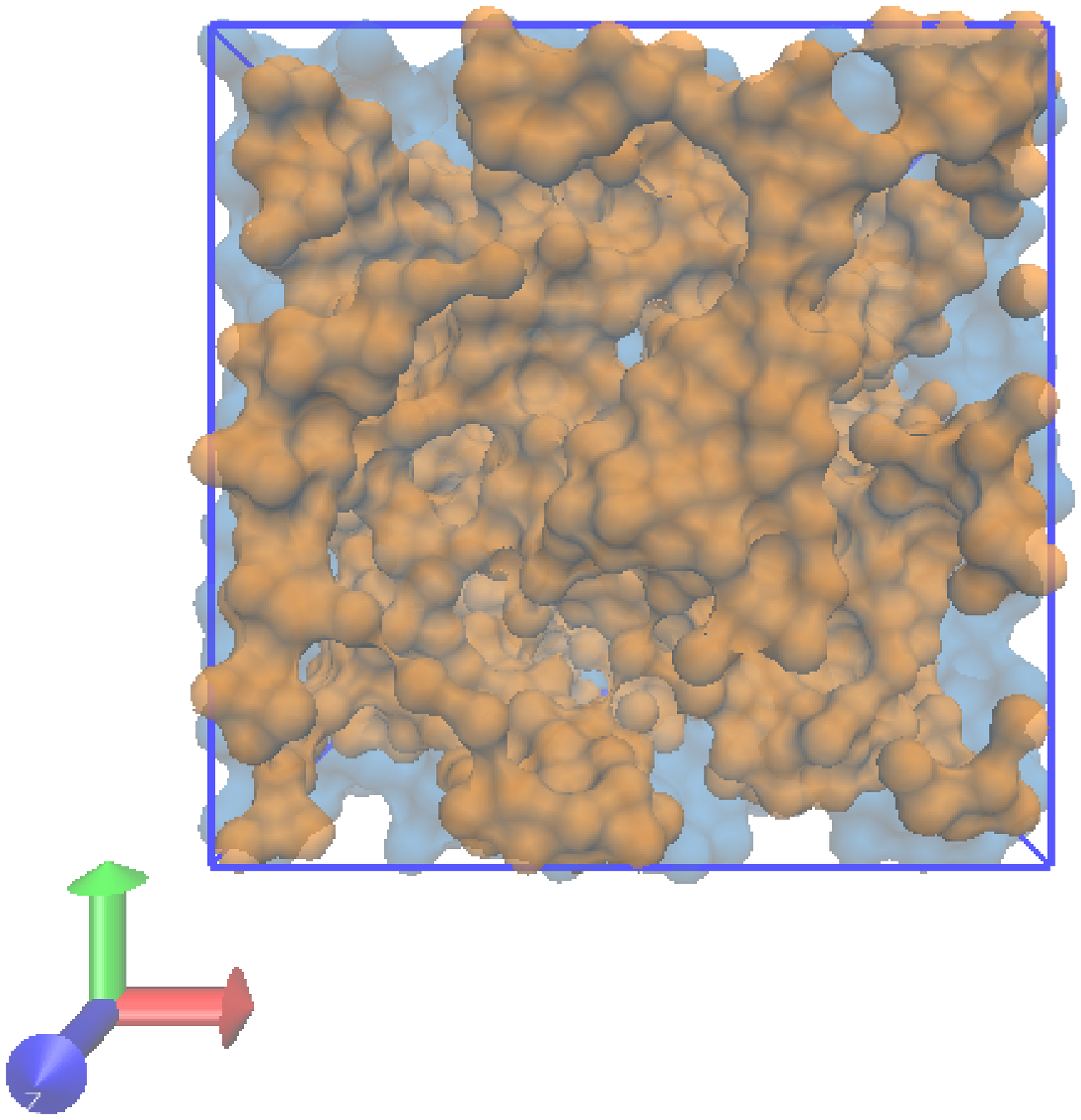}
\label{snaphots_x03_semi_07}}\\
\caption[width=0.47\columnwidth]{Surface plots for protons (orange) and 
neutrons (blue) at  $T=0.7\,$MeV ($\rho=0.05$ and $x=0.3$). The left snapshot 
shows a view from one side of the cell (in full colors). The right snapshot 
shows the same view in semi-transparent colors.  } 
\label{snaphots_x03}
\end{figure}

According to Fig.~\ref{clusters_rho05_03_500}, a few clusters exist for 
``cold'' 
neutron star matter. These get gradually connected as the system is warmed, 
until no clusters can be distinguished at all. The connecting process appears 
to be fulfilled near $T=1\,$MeV (notice the vanishing slope in 
Fig.~\ref{clusters_rho05_03_500}). Thus, the lack of visible energy ``jumps'' 
in Fig.~\ref{heating_energy_3} for the proton ratio $x=0.3$ does not mean the 
absence of a \textit{pasta} breakdown, but the embedding of this topological 
transition into the spread around neutrons. Fig.~\ref{snaphots_x03} pictures 
the 
topology transition in full colors and semi-transparent colors for 
$T\simeq0.7\,$MeV. \\  

\subsubsection{\label{thermal_conductivity_asymmetric} The thermal conductivity 
$\kappa$}

\begin{figure}[htbp!]
\includegraphics[width=\columnwidth]
{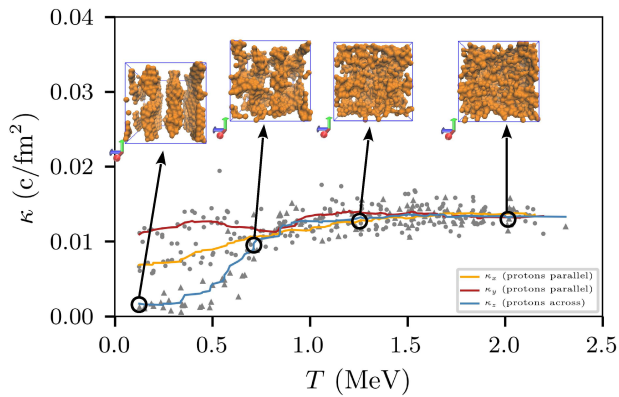}
\caption{\label{kappa_protons_asymmetric_snapshots} (On-line color only) Proton 
thermal conductivity $\kappa$ vs. temperature for $\rho=0.05$ and proton ratio 
$x=0.3$. The smoothening was done following a moving average procedure of 
$\pm10$ points. The rounded gray points correspond to data obtained along the 
lasagna direction. The triangular gray points correspond to data across the 
lasagna structure.  } 
\end{figure}

The thermal conductivity for non-symmetric neutron star matter was computed in 
the same way as in Section~\ref{thermal_conductivity}. 
Fig.~\ref{kappa_protons_asymmetric_snapshots} shows the proton thermal 
conductivity behavior for the $x=0.3$ situation, evolving from ``cold'' (solid) 
temperatures to ``warm'' ones. The corresponding snapshots (protons only) are 
also exhibited. \\

Notice that the (qualitative) patterns appearing in 
Fig.~\ref{kappa_protons_asymmetric_snapshots} resemble those exhibited in 
Fig.~\ref{kappa_all_vs_protons} for the symmetric situation (and for similar 
density). The proton conductivity across the slabs vanishes. Besides, the 
conductivity ``decoupling'' is present on either symmetric and non-symmetric 
matter, in correspondence with the topological transformations (\textit{i.e.} 
the cluster drop-off shown in Fig.~\ref{clusters_rho05_05_400} and 
Fig.~\ref{clusters_rho05_03_500}, respectively). The ``decoupling threshold'' 
at 
this instance, however, appears somewhat biased with respect to the symmetric 
situation.  \\

\begin{figure}[htbp!]
\includegraphics[width=\columnwidth]
{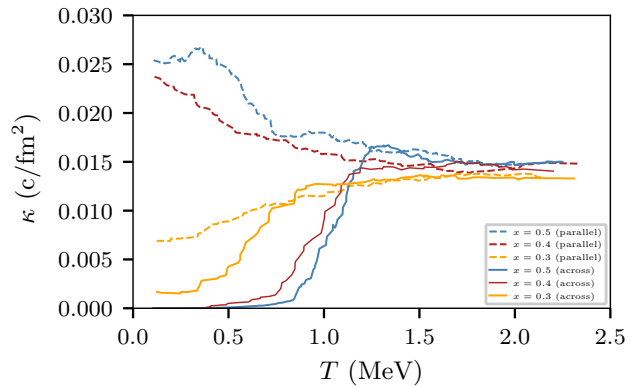}
\caption{\label{kappa_protons_asymmetric} (On-line color only) Proton thermal 
conductivity $\kappa$ vs. temperature for $\rho=0.05$ and proton ratios in the 
range $x=0.3$ to 0.5 (see insert for details). The smoothening was done 
following a moving average procedure of $\pm10$ points. The dashed lines 
correspond to the thermal conductivity values along (parallel) the pasta 
structure. The continuous lines correspond to the thermal conductivity across 
the pasta structure.  } 
\end{figure}

Fig.~\ref{kappa_protons_asymmetric} brings out the complete picture for the 
proton thermal conductivity $\kappa$. Although the profiles are qualitatively 
similar, the asymmetric proton conductivity values scale down with respect to 
the symmetric proton conductivity. Say, the parallel conductivity for $x=0.3$ 
(see Fig.~\ref{kappa_protons_asymmetric}) never surpasses $0.01\,$c/fm$^{2}$, 
while the corresponding values for $x=0.5$ appear always above. 
Furthermore, the later reports a maximum at the solid state (``cold'' 
temperatures), while the former does not.   \\

\begin{figure}[htbp!]
\includegraphics[width=\columnwidth]
{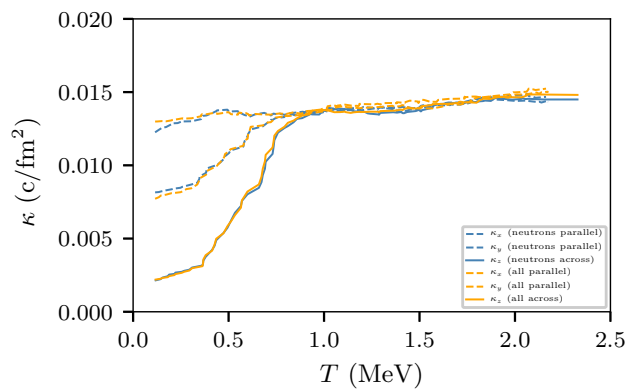}
\caption{\label{kappa_neutrons_all_asymmetric} (On-line color only) Thermal 
conductivity $\kappa$ vs. temperature for $\rho=0.05$ and proton ratio 
$x=0.3$. The blue lines correspond to the neutron conductivity, while the 
orange 
lines consider \textit{all} the nucleons (see insert for details). The dashed 
lines (regardless for the color) correspond to the conductivity along 
(parallel) the spaghetti structure. The continuous lines, instead, correspond 
to 
any direction across the spaghetti structure. The smoothening was done 
following 
a moving average procedure of $\pm10$ points.  } 
\end{figure}

The overall thermal conductivity (that is, considering \textit{all} the 
nucleons) appears to be very similar for symmetric and non-symmetric matter, at 
``warm'' temperatures. The corresponding profile for $x=0.3$ is shown in 
Fig.~\ref{kappa_neutrons_all_asymmetric}. The neutron thermal conductivity is 
also included. Both profiles are remarkably similar, meaning that the thermal 
conduction for $x=0.3$ is mostly achieved by neutrons. \\

The above observations indicate that the thermal conductivity for non-symmetric 
matter shares the same qualitative behavior as the symmetric matter, despite 
that \textit{pastas} are now embedded in a cloud of neutrons. The neutron  
thermal conductivity, though, resembles better the overall conductivity than 
the proton conductivity. \\

\section{\label{conclusions}Conclusions}

The thermal conductivity of \textit{pastas} raises as a complex magnitude that 
is far from attaining a well established behavior. Researchers admit that 
variations of $\kappa$ of (at least) an order of magnitude can be expected at  
sub-saturation densities, and temperatures below 2~MeV. The proton fraction is 
also a significant source of variations and a challenging field of 
investigation.  \\   

Since the \textit{pasta} structures may become too complex for an increasing 
number of nucleons, we focused on simple structures (say, \textit{lasagnas} and 
\textit{spaghettis}) housing 4000 nucleons. We arrived to the main conclusion 
that the \textit{pasta} breakdown process accomplishes a dramatic change of the 
phononic thermal conductivity. Neutron star matter switches 
from a strong non-isotropic regime (at the well-formed \textit{pastas} regime) 
to an isotropic one, as temperature increases. This occurs sharply around  
$T\simeq 1\,$MeV for symmetric matter, and somewhat below this threshold for 
non-symmetric matter. \\

The above conclusion is a compelling reason for associating the \textit{pasta} 
topological transition to low (or high) phononic thermal 
conductivities, although an estimate of the ``effective'' $\kappa$ across the 
neutron star crust is not yet available. The \textit{pasta} breakdown 
threshold, 
though, appears as a key issue for this estimate.  \\

An insight into the breakdown threshold shows the crucial role of the
``connecting'' nucleons between \textit{lasagnas} or \textit{spaghettis}. The 
heat flux at the threshold can only be sustained if a (small) fraction of 
nucleons bridge  the slabs or rods. But as soon as these bridges open, the 
thermal conductivity (in the bridging direction) drops off. The temporary 
character of these connections leads to significant fluctuation of $\kappa$ 
close to the threshold. \\

The phononic thermal conductivity appears ``decoupled'' at the 
well-formed \textit{pasta} regime (say, below $T\simeq 1\,$MeV for symmetric 
matter). The heat flux drops off across void regions, but enhances along the 
\textit{pasta} structure. $\kappa$ exhibits a maximum at the solid-liquid 
transition for symmetric matter. We may  expect, therefore, that the coldest 
pathways in the crust become also true directions for heat conduction. \\

Attention was claimed in the literature on the existence of neutron-rich 
layers in the deep crust. The (overall) phononic thermal 
conductivity does not exhibit relevant variations with respect to the proton 
fraction at such ``warm'' temperatures. The heat conduction, however, appears 
to be driven by the richest species. Recall that \textit{pastas} are actually 
not present above $T\simeq 1\,$MeV.  \\

The ``decoupling'' phenomenon still appears in the well-formed \textit{pasta} 
regime for neutron-rich matter. The \textit{pastas} are actually embedded in a 
cloud of neutrons, seemingly isolated since the thermal conductivity drops off 
across the structures. Besides, the solid-liquid transition does not accomplish 
a conductivity enhancement, as occurs in symmetric matter. The irregularity of 
the slabs (due to the low proton fraction) seems to be the possible reason for 
this instance. We support this conclusion on the fact that the neutron thermal 
conductivity level for $x=0.3$ (at the solid-liquid transition) is similar to 
the proton conductivity for $x=0.5$ and $\rho=0.03-0.04$. That is, both 
situations account for the same topological ``defects'', although for two 
different conditions. \\

\begin{acknowledgments}
This work was supported by the National Scientific and Technical 
Research Council, Argentina (spanish: Consejo Nacional de Investigaciones 
Cient\'\i ficas 
y T\'ecnicas - CONICET) grant number PIP 2015-2017 GI, founding 
D4247(12-22-2016). AS was partially supported by  U.S. National Science 
Foundation 
under contract: CBET 1404823. 
\end{acknowledgments}

\newpage 
\bibliography{paper}

\end{document}